\documentstyle[fleqn]{tp}

\input epsf.sty
\input psfig.sty

\def\etal{{\it et\thinspace al\/}}

\def\kms{\mbox{km s$^{-1}$}}
\def\mpc{\mbox{Mpc}}

\def\ltsima{$\; \buildrel < \over \sim \;$}
\def\simlt{\lower.5ex\hbox{\ltsima}} 
\def\gtsima{$\; \buildrel > \over \sim \;$} 
\def\simgt{\lower.5ex\hbox{\gtsima}}

\def\etal{{\rm et al.~\/}} 
\def\kms{\ifmmode {\rm \ km \ s^{-1}}\else$\rm km s^{-1}$\fi} 
\def\mpc{$\ {h^{-1}\rm Mpc}$}

\def\si{{\sigma_{8,{\rm iras}}}} 
\def\s8{{\sigma_8}} 
\def\h{{h}}
\def\ns{{n_{\rm s}}} 
\def\alphacmb{{\vec{\alpha}}_{\rm cmb}}
\def\alphairas{{\vec{\alpha}}_{\rm iras}}
\def\alphajoint{{\vec{\alpha}}_{\rm joint}}
\def\omegam{{\Omega_{\rm m}}}
\def\omegacdm{{\Omega_{\rm cdm}}} 
\def\omegab{{\Omega_{\rm b}}}
 
\def\omegal{{\Omega_\Lambda}}

\def\ns{{n_{\rm s}}} 
\def\omegabh2{{\omegab h^2}} 
\def\bi{{b_{\rm iras}}}
\def\betai{{\beta_{\rm iras}}}

\def\Mpc{\, h^{-1} \, {\rm Mpc}}

\begin{document}

\twocolumn[
\title{Joint Cosmological Inference from galaxy surveys, the Cosmic Microwave Background 
and the 
X-Ray Background}
\author{Ofer Lahav$^{1,2}$ and Sarah L. Bridle$^{3}$\\
{\it $^1$Racah Institute of Physics, The Hebrew University, Jerusalem 
91904, Israel}\\
{\it $^2$Institute of Astronomy, Madingley Road, Cambridge CB3 0HA, UK}\\
{\it $^3$Astrophysics Group, Cavendish Laboratory, Madingley Road,Cambridge CB3 0HE, UK}}
\vspace*{16pt}

ABSTRACT.\
We discuss cosmological inference from galaxy surveys, the X-Ray Background
(XRB) 
and the Cosmic Microwave Background (CMB).
We assume a family of Cold Dark Matter (CDM) models
in a spatially flat universe with an initially 
scale-invariant spectrum and a cosmological constant.
Joint analysis of the CMB and IRAS resdhift surveys yields as
optimal parameters (for fixed $\Omega_b h^2 = 0.024$):
 
$\omegam = 1- \omegal = 0.4$, $h = 0.53$, $Q = 17$ $\mu$K, and $\bi = 1.2$.
For the above parameters the normalisation and 
shape of the mass power-spectrum 
are $\sigma_8=0.7$ and $\Gamma=0.15$, 
and the age of the Universe is $16.5$~Gyr.
Assuming a standard CDM model, the joint IRAS and XRB analysis 
gives bias parameters of $\bi = 1.1$ and $b_x(z=0) = 2.6$.
When combining CMB and IRAS with XRB data we  show that
standard CDM cannot fit all  three data sets. 
We also use these data sets to 
verify the Cosmological Principle and to 
constrain the fractal dimension of the universe 
on large scales.
\endabstract]

\markboth{O. Lahav and S.L. Bridle}{Joint cosmological inference}

\small

\section{Introduction}

Observations of large scale structure (LSS) and the Cosmic Microwave
Background (CMB) each place separate constraints on the values of
cosmological parameters.

Estimates derived separately from each of these two data sets have
problems with parameter degeneracy. In the analysis of LSS data, there
is uncertainty as to how well the observed light distribution traces
the underlying mass distribution. The light-to-mass linear bias, $b$,
introduced to account for this uncertainty, affects the value of many
central cosmological parameters, and makes any identified optimum
degenerate. Similarly, on the basis of CMB data alone,
there is considerable degeneracy between $\h = H_0 /
100 \kms {\rm Mpc}^{-1}$  and the energy density $\omegal$ due to the
cosmological constant. This leads to poor estimation of
the baryon ($\omegab$) and total mass ($\omegam$) densities.

Several authors (e.g. Gawiser \& Silk 1998; Eisenstein, Hu \& Tegmark 1998) 
have recently discussed joint analysis of cosmological probes. 
Here we summarize results from 
Webster et al. (1998) and Bridle et al. (in preparation) 
which combine CMB, XRB  and IRAS data. 
We present a self-consistent formulation of CMB and LSS parameter
estimation. In particular, our method expresses the effects of the
underlying mass distribution on both the CMB potential fluctuations
and the IRAS redshift distortion. 
The clustering
of galaxies in redshift-space is systematically different from that in
real-space. The mapping between the two is a
function of the underlying mass distribution, in which the galaxies
are not only tracers, but also velocity test particles.
The joint analysis  breaks the degeneracy inherent
in an isolated analysis of either data set, and places tight
constraints on several cosmological parameters. 
For simplicity, we restrict our attention to inflationary,
Cold Dark Matter (CDM) models, assuming a flat universe with linear,
scale-independent biasing.


\section{The CMB}

The compilation of CMB data set is described in Hancock et al. (1998) and 
in Webster et al. (1998) and is shown in Figure 1.
\begin{figure*}
\centering\mbox{\psfig{figure=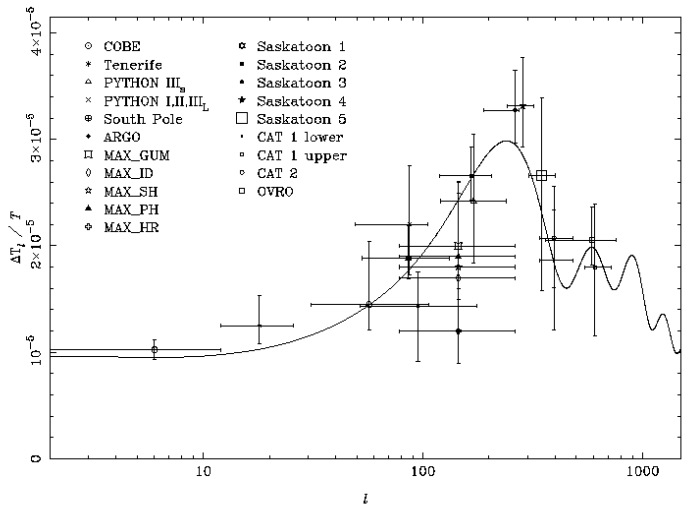,height=8cm}}
\caption[]{ A compilation of CMB measurements and the `best fit' universe based
on the joint CMB+IRAS estimation,  for parameters
given in Table 1.}
\label{fig:1}
\end{figure*}
As in Hancock et al. (1998)
we form a chi-squared $\chi^2(\alphacmb)$ statistic between the flat
bandpowers $\Delta T_l$ (20 measurements)
for a given set of parameter values $(\alphacmb)$. 
Since the CMB data points were chosen such that no 
two bandpower estimates come from experiments which observed
overlapping patches of sky and had overlapping window functions, we
may consider them as {\em independent} estimates of the CMB power
spectrum.
As the cosmic variance has already been  taken into account in deriving 
the flat bandpower estimates, the likelihood function 
is given simply by
${\cal L}_{\rm cmb} \propto e^{-\chi^2/2}$.

We assume that the Universe is spatially flat, and
that there are no tensor contributions to the CMB power spectrum. We
take the primordial scalar perturbations to be described by the
Harrison-Zel'dovich power spectrum for which $\ns=1$, and further
assume that the optical depth to the last scattering surface is zero.

The normalisation of the CMB power spectrum is determined by $Q$,
which gives the strength of the quadrupole in $\mu K$.
Hereafter
$\omegacdm$ and $\omegab$ denote the density of the Universe in
CDM and the baryons respectively, each in units of the
critical density. Given that we assume a flat universe, but
investigate models where
$ \omegam \equiv (\omegacdm + \omegab) < 1 \ $,
the shortfall is made up through a non-zero cosmological constant
$\Lambda$ such that
$\omegal = 1 - \omegam =  {{\Lambda}/({3{H_0^2}})} \ $ .
Furthermore, we restrict our attention to models that satisfy the
nucleosynthesis constraint $\Omega_b h^2 = 0.024$ (Tytler et al. 1996).
Thus we consider the reduced set of CMB parameters
\begin{equation}
\alphacmb \equiv \{Q, h, \omegacdm\} \ .
\end{equation}
%


\section{IRAS}


We use the
1.2 Jy IRAS survey (Fisher et al. 1995),  consisting of 5313 galaxies,
covering 87.6\% of the sky.


Here we assume linear, scale-independent biasing, where
$\bi$ measures the ratio between fluctuations in the IRAS galaxy
distribution and the underlying mass density field:
\begin{equation}
{\left({{\delta\rho}/{\rho}}\right)}_{\rm iras} =
\bi\,{\left({{\delta\rho}/{\rho}}\right)_{\rm m}} \ .
\end{equation}
We note that biasing may be 
non-linear, stochastic, non local, scale dependent, epoch dependent 
and type dependent (e.g.  Dekel \& Lahav 1998, 
Tegmark \& Peebles 1998, Blanton et al. 1998,  and references therein).
For a linear  bias parameter, $\bi$,
the velocity and density fields in linear theory 
are linked by a proportionality factor 
$\betai \equiv {{\omegam^{0.6}}/{\bi}}$.

Statistically, the fluctuations in the real-space galaxy distribution
can be described by a power spectrum, ${P(k)}$, which is
determined by the {\it{rms}} variance in the observed galaxy field,
measured for an 8\mpc\ radius sphere ($\si$) and a shape parameter
(e.g.  $\Gamma$ below). The observed $\si$ is
related to the underlying $\s8$ for mass through the bias parameter,
such that $\si = \bi\,\s8$.

We follow the spherical harmonic approach of   Fisher, Scharf \& Lahav
(1994).
The likelihood of the survey harmonics can be
calculated as
\begin{equation}
{\cal L}_{\rm iras} \propto  {{\left| {\bf A} \right|}^{-{{1}\over{2}}}}\,
{ {\rm exp}^{ \left(
{{-{{1}\over{2}}}\left[{{\vec{a}}^{\rm T}}\,{\bf A}^{-1}\,{\vec{a}}\right]}
\right)}} \ .
\label{iraslike}
\end{equation}
Here $\vec{a}$ is the vector of observed harmonics for different
radial shells 
${\bf A}$ is the corresponding covariance matrix, which
depends on the predicted harmonics (including shot noise). 
Note that the argument of the exponent in equation~\ref{iraslike}
is simply 
$(-\chi^2/2)$, and that here the normalisation of the likelihood
function does depend on the free parameters (unlike in the CMB 
likelihood function).

Since our analysis is
valid only in the linear regime, we restrict the likelihood
computation to $l_{\rm max} = 10$ (corresponding to 120 degrees 
of freedom). 
The IRAS window functions are sensitive to scales 
$k \sim 0.01-0.1 $.
To summarize, the IRAS likelihood function
has a parameter vector 
\begin{equation}
\alphairas \equiv \{\betai, \si, \Gamma\} \ .
\end{equation}
%


\section{Joint analysis CMB+IRAS}

Given the large number of parameters available between the two models,
it is important both to find links for joint optimisation, and to
decide which parameters can be frozen. From section 3, we have six
variables between the two models: \{$Q$, $\h$, $\omegacdm$,
$\betai$, $\si$, $\Gamma$\}. These can be reduced further by
expression in terms of core cosmological parameters. The IRAS
normalisation can be calculated as $\s8 \equiv f(\omegam, Q,
\Gamma)$, while the CDM shape parameter $\Gamma$ depends on 
$\omegam\,\h$ and $\omegab$ (Sugiyama 1995). 
On the other hand, 
$\omegam = \omegacdm + \omegab$, $\betai = \Omega_{\rm m}^{0.6}/\bi$, 
and $\si = \s8\bi$. Hence, the final, joint parameter space is
\begin{equation}
\alphajoint \equiv \{\h, Q, \omegam, \bi\}\ .
\label{alphajoint}
\end{equation}
As the IRAS and CMB probe very different scales and hence are 
assumed to be uncorrelated,  
the joint likelihood is given by
\begin{equation}
{\rm ln}\left({\cal L}_{\rm joint}\right) = 
{\rm ln}\left({\cal L}_{\rm cmb}\right) + 
{\rm ln}\left({\cal L}_{\rm iras}\right) \ .
\label{jointlike}
\end{equation}
%


The joint likelihood (equation~\ref{jointlike}) was maximized with
respect to the 4 free parameters (equation~\ref{alphajoint}) 
and the best fit
parameters are shown in Table~1.
%
\vskip0.2cm
\centerline{\vbox{
{\footnotesize
{\bf Table~1. --- Parameter values at the joint optimum.}
The 68\% confidence limits are shown,
calculated for each parameter by marginalising the likelihood over the other 
variables.\label{valuetable}}\\\\
\begin{tabular}{@{}lc r@{.}l @{$\,<\,$} c @{$\,<\,$} r@{.}l}
$\omegam$    &$0.39$    &  $0$ & $29$ &$\omegam$&$0$&$53$\\
$h$          &$0.53$    &  $0$ & $39$ &$h$      &$0$&$58$ \\
$Q$ ($\mu$K) &$16.95$   &  $15$& $34$ &$Q$      &$17$&$60$\\
$\bi$        &$1.21$    &  $0$ & $98$ &$\bi$    &$1$&$56$\\
\\
\end{tabular}
}}
%
It follows that  $\Omega_b=0.085$, $\s8=0.67$, $\sigma_{8,{\rm iras}}=0.81$,
$\Gamma=0.15$, $\betai=0.47$ and the age of the Universe is $16.5$
Gyr. For this set of parameters, we find the values of the reduced $\chi^2$
for the IRAS and CMB data respectively to be 1.18 and 1.03, confirming
that both data-sets agree well with the models used.  Taking the CMB
and IRAS data together the total reduced $\chi^2$ is 1.16. 

We  note that our joint IRAS \& CMB 
optimal values for $\sigma_{8,{\rm iras}}, \beta_{{\rm iras}}$
and $\Gamma$ (Table~1) are in perfect agreement with 
the values derived from IRAS alone 
(Fisher et al. 1994, Fisher 1994).
However at fixed $\sigma_{8, {\rm iras}} = 0.69$ 
based on the IRAS correlation function Fisher et al. (1994) found 
a higher $\beta_{{\rm iras} } = 0.94 \pm 0.17 $
and $\Gamma = 0.17 \pm 0.05 (1-\sigma)$.

\begin{figure*}
\centering
{\psfig{file=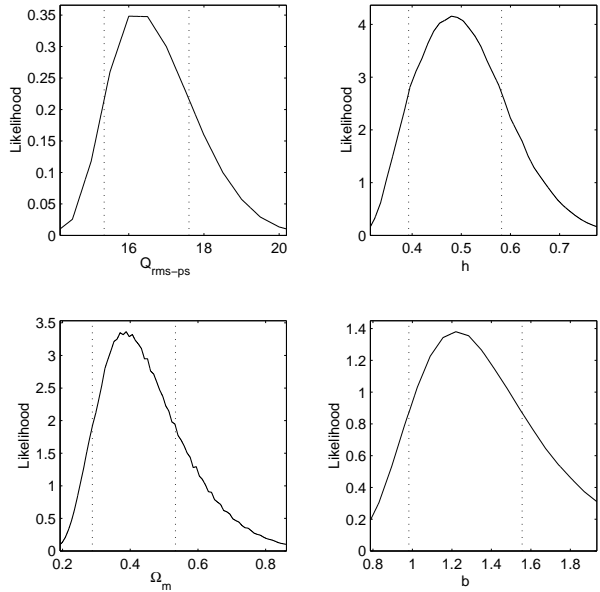,height=8cm}}
\caption
{The one-dimensional marginalised probability distributions for
each of the four parameters. The vertical dashed lines denote the 68\%
confidence limits. The horizontal plot limits are at the 99\%
confidence limits.} 
\label{margplots}
\end{figure*}

To obtain 68 per cent confidence limits on each of the free parameters
it is necessary to marginalise over the remaining free parameters. 
The marginalised distribution for each parameter is
shown in Figure~\ref{margplots}, in which the dashed vertical lines denote
the 68\% confidence limits quoted in Table~1.

In addition 
we evaluated the covariance matrix at the joint optimum. 
The most strongly correlated parameters are 
$\omegam$ and $h$, with (normalized) correlation coefficient of $(-0.82)$ .


\section{The XRB}

 Although discovered in 1962, the origin of
 the X-ray Background (XRB) is still unknown,  
 but is likely
 to be due to sources at high redshift 
 (for a review see Fabian \& Barcons 1992).
 Here we shall not attempt to speculate on the nature of the XRB sources.
 Instead, we {\it utilise} the XRB as a probe of the density fluctuations at
 high redshift.  The XRB sources are probably
 located at redshift $z < 5$, making them convenient tracers of the mass
 distribution on scales intermediate between those in the CMB as probed
 by COBE, and those probed by optical and IRAS redshift
 surveys (see Figure 3).

The interpretation of the results depends somewhat on the nature of
the X-ray sources and their evolution.  The rms dipole and higher
moments of spherical harmonics can be predicted (Lahav, Piran \& Treyer 
1997) in the
framework of growth of structure by gravitational instability from
initial density fluctuations.
By comparing
the predicted multipoles to those observed by HEAO1 
(Treyer et al. 1998)
we estimate the amplitude of fluctuations for an
assumed shape of the density fluctuations 
(e.g. CDM models).  
Figure 3 shows the amplitude of fluctuations derived at the 
effective scale $\lambda \sim 600 h^{-1}$ Mpc probed by the XRB. 
The observed fluctuations in the XRB
are roughly as expected from interpolating between the
local galaxy surveys and the COBE CMB experiment.
The rms fluctuations 
${ {\delta \rho} \over {\rho} }$
on a scale of $\sim 600 h^{-1}$Mpc 
are less than 0.2 \%.

\begin{figure*}
\protect\centerline{
\psfig{figure=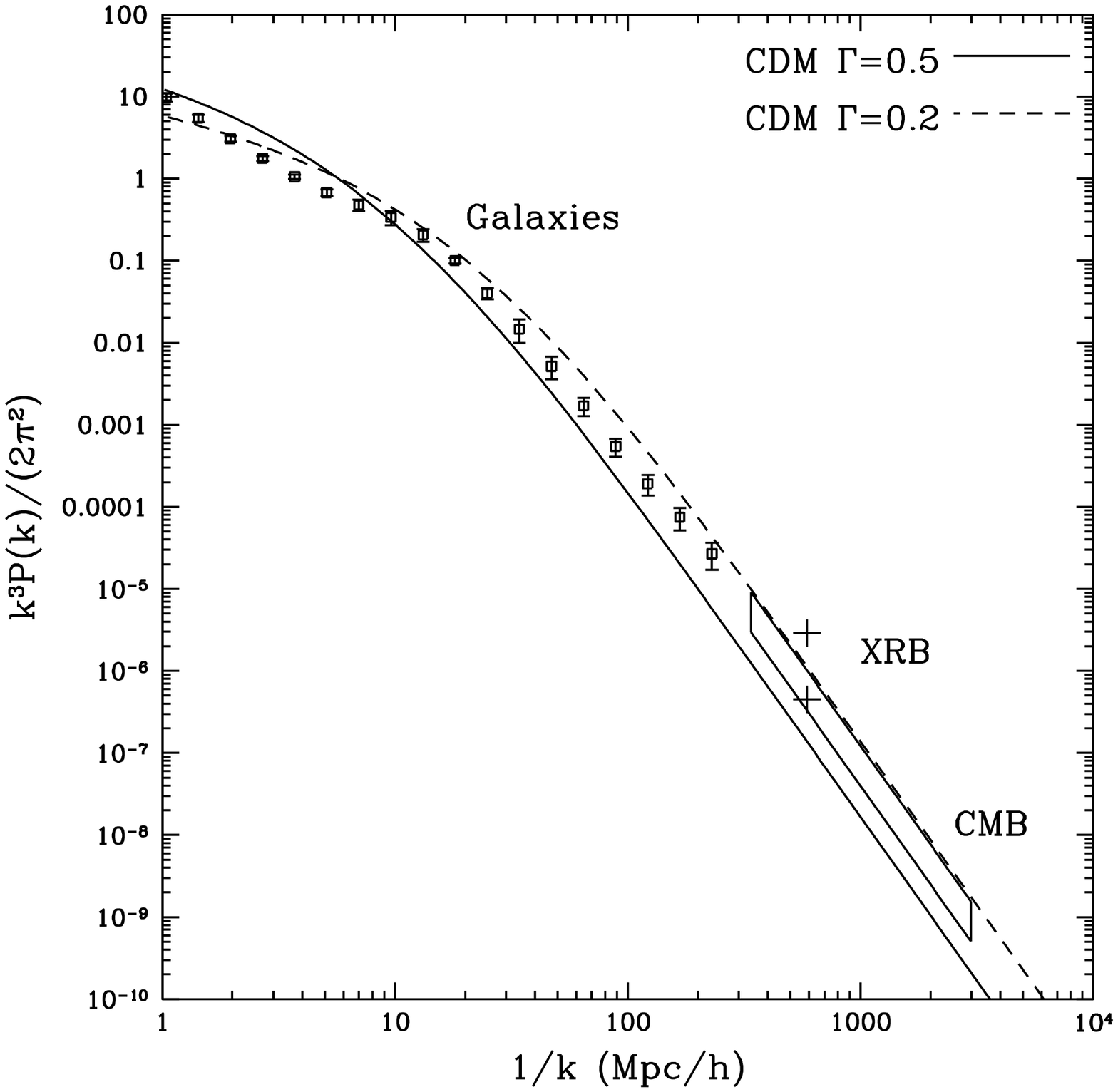,height=3truein,width=3truein}}
\caption[]{
A compilation of rms density fluctuations, 
$({ {\delta \rho} \over {\rho} })^2 \sim k^3 P(k)$,
on different scales 
from various observations: a galaxy survey, the X-ray
Background and Cosmic Microwave Background experiments.  
The crosses represent  constraints from 
the XRB HEAO1 quadrupole (Lahav et al. 1997, Treyer et al. 1998).
The top and bottom crosses are estimates of the amplitude of 
the power-spectrum at $k^{-1} \sim 600 h^{-1}$ Mpc,
assuming CDM power-spectra with shape parameters
$\Gamma=0.2$ and $0.5$ respectively, and an Einstein-de Sitter
universe.  The fractional error on the XRB amplitudes (due to the
shot-noise of the X-ray sources) is about 30\%.
The solid and dashed lines correspond to the standard CDM
power-spectrum (with shape parameter $\Gamma = 0.5$) and a
`low-density' CDM power-spectrum (with $\Gamma=0.2$), respectively,
assuming $\sigma_8=1$ in both cases.
The open squares at small scales are estimates of the
power-spectrum from 3D inversion of the angular APM galaxy
catalogue (Baugh \& Efstathiou 1994). 
The elongated box at large scales
represent the COBE 4-yr  CMB measurement.
The COBE box corresponds to a quadrupole Q=18.0 $\mu K$ for a
Harrison-Zeldovich mass power-spectrum, via the Sachs-Wolfe 
effect, or $\sigma_8 =1.4$ 
for a standard CDM model (Gawiser and Silk 1998).}  
\end{figure*}

\subsection{Joint IRAS+XRB estimation} 

For simplicity we restricted the analysis to the standard CDM model
($\omegam=1, h=0.5$) with normalization $\sigma_8 = 0.7$.
Here we assumed a revised 
$\omegab h^2 = 0.019$ (Burles \& Tytler 1998).
As the sources of the XRB cover a wide range in redshift we assumed
an epoch-dependent biasing  of the form (Fry 1996):
\begin{equation}
b_x(z) = b_x(0) + z [b_x(0) -1 ]
\end{equation}
and that the X-ray emissivity varies like 
$(1+z)^{2.6}$ out to $z_{max} =6.4$ (Treyer et al. 1998).
In the likelihood analysis  we used the HEAO1 data with 
a Galactic mask, and harmonics $1 \leq l \leq 10$. 
We then maximized the joint likelihood IRAS+XRB (assumed to be 
uncorrelated) with respect to 
2 free biasing parameters, and found 
at the $\bi = 1.1  $ and $b_x(0) = 2.6 $.
The goodness-of-fit is $\chi^2_{iras} = 1.22$ and   
$\chi^2_{xrb} = 1.01$.

\subsection{Joint IRAS+XRB+CMB estimation} 

We then  added the CMB and solved for 3 free parameters: 
$\bi = 0.7$,  $b_x(0) = 1.8 $ and $\sigma_8 = 1.0$.
While the $\chi^2$ for IRAS and XRB are acceptable, 
the CMB fit is very poor, $\chi^2_{cmb} = 2.7$. 
Hence standard CDM cannot fit these 3 data sets simultaneously.
We intend to generalise this analysis for other models.

\section{Comparison with other studies}

The results of the CMB+IRAS optimisation are in reasonable agreement
with other current estimates. 
The relatively low value of $\omegam \approx 0.4$ is close to that found by
others (White et al. 1993, Bahcall et al. 1997), and is in line with
recent supernovae results (Perlmutter et al. 1998). However, given 
the assumption of a
flat universe, it requires a very high cosmological constant 
($\omegal = 0.6$). Gravitational lensing measurements 
have constrained $\omegal < 0.7$ (Kochanek 1996).
Our value for the Hubble constant, 
$\h = 0.53$, agrees well with several other measurements
(Sugiyama 1995, Lineweaver et al. 1997), but
falls at the low end of the generally accepted range from local
measurements (Freedman et al. 1998). Assuming the nucleosynthesis
constraint $\Omega_b h^2 = 0.024$ 
the optimal baryon density is found to be $\omegab = 0.085$.
Our value for the combination $\s8 \omegam^{0.6} = 0.38$ 
is lower than the one derived from 
measurements from the peculiar velocity field, 
$\sigma_8 \omegam^{0.6} \approx 0.8$ (Freudling et al. 1998).
Our values are closer to the combination
derived from cluster abundance  
$\sigma_8 \omegam^{0.5} \approx 0.5 $ (Eke et al. 1998).
Finally, for spatially-flat universes the time since
the Big Bang 
for the values of our $\omegam$ and $h$ 
at the joint optimum is
$16.5$ Gyr. 

On the IRAS side, $\betai = 0.47$ is in agreement with several other
measurements (Willick et al. 1997), although there are
other measurements  which place $\betai$ much higher
(Sigad et al. 1997). 
Finally, the IRAS mass-to-light
bias is seen to be slightly greater than unity ($\bi = 1.2$), suggesting
that the IRAS galaxies (mainly spirals) are reasonable (but not perfect) 
tracers of the underlying mass distribution.
On the other hand, the XRB sources are strongly biased
relative to the mass ($b_x(0) = 2.6$).

  \section {Is the FRW Metric Valid on Large Scales ?}

The Cosmological Principle was first adopted when observational
cosmology was in its infancy; it was then little more than a
conjecture. Observations could not then probe to significant redshifts, the
`dark matter' problem was not well-established and the 
CMB and the XRB were still unknown.  
If the  Cosmological Principle turned out to be invalid 
then the consequences to our understanding of cosmology would be dramatic, 
for example the conventional way of interpreting the age of the universe, 
its geometry and matter content would have to be revised. 
Therefore it is 
important to revisit this underlying assumption in the light of new
galaxy surveys and measurements of the XRB and CMB.
The question of whether the universe 
is isotropic and homogeneous on large scales
can also be  phrased in terms of the fractal structure of the 
universe.
A fractal is a geometric shape that is not homogeneous, 
yet preserves the property that each part is a reduced-scale
version of the whole.
If the matter in the universe were actually 
distributed like a pure fractal on all scales then the 
Cosmological Principle 
would be invalid, and the standard model in trouble.
As shown in Figure 3 
current data already strongly constrain any non-uniformities in the 
galaxy distribution (as well as the overall mass distribution) 
on scales $> 300 \Mpc$.

If we count, for each galaxy,
the number of galaxies within a distance $R$ from it, and call the
average number obtained $N(<R)$, then the distribution is said to be a
fractal of correlation dimension $D_2$ 
if $N(<R)\propto R^{D_2}$. Of course $D_2$
may be 3, in which case the distribution is homogeneous rather than
fractal.  In the pure fractal model this power law holds for all
scales of $R$.

The fractal proponents (Pietronero et al. 1997)  have
estimated $D_2\approx 2$ for all scales up to $\sim 1000\Mpc$, whereas
other
groups 
have obtained scale-dependent values 
(for review see Wu, Lahav \& Rees 1998,  and references therein).

If we  assume
homogeneity on large scales, 
then we have a direct mapping
between  correlation function $\xi(r)$ (or the Power-spectrum)  
and $D_2$. For 
$\xi(r) \propto r^{-\gamma}$
it follows that $D_2=3-\gamma$ if $\xi\gg 1$, while
if $\xi(r)=0$ then $D_2=3$. 
We note that it is
inappropriate to quote a single crossover scale to homogeneity, for
the transition is gradual.

Direct estimates of $D_2$ are not possible for much larger scales, but
we can calculate values of $D_2$ at the scales probed by the XRB and
CMB by using CDM models normalised with the XRB and CMB as described
above.  The resulting values 
are consistent with $D_2=3$ to within
$10^{-4}$  from the XRB on scales $\sim 500 \Mpc$ 
and $2 \times 10^{-5}$ from the CMB 
on $\sim 1000 \Mpc$ (Wu et al. 1998).
Isotropy does not imply homogeneity, but the near-isotropy of the CMB
can be combined with the Copernican principle that we are not in a
preferred position.

\section{Discussion} 
The near future will see a dramatic increase in LSS data
(e.g.\ the PSCZ, SDSS, 2dF and 6dF surveys) and detailed measurements of 
the CMB fluctuations on sub-degree scales (e.g.\ from the Planck
Surveyor and MAP satellites). 
These will allow more accurate parameter estimation
and exploration of a wider range of models.
In particular 
we emphasise that the naive linear biasing 
should be generalised to more realistic scenarios. 
Other cosmological probes (e.g. Supernovae, clusters, peculiar velocities
and radio sources) can be added to the analysis 
to set tighter constraints on the parameter space.


\subsection*{Acknowledgments}
We thank M. Hobson, A. Lasenby, M. Rees, G. Rocha , C. Scharf, 
M. Treyer, M. Webster, and K. Wu 
for their contribution for the  work presented here.



\label{lastpage}

\end{document}